\documentclass[12pt]{article}
\usepackage{graphicx}
\usepackage{bm}        
\usepackage{amssymb}   
\usepackage{hyphenat}

\input epsf
\def\beq{\begin{eqnarray}}
\def\eeq{\end{eqnarray}}

\def\ba{\begin{eqnarray}}
\def\ea{\end{eqnarray}}
\def\beq{\begin{eqnarray}}
\def\eeq{\end{eqnarray}}

\def\p{{\cal P}}

\def\L*{{\cal L}_*}
\def\L{\mathcal{L}}
\def\({\left(}
\def\){\right)}

\def\nn{\nonumber}
\def\p{\partial}
\def\mn{_{\mu \nu}}

\def\p{\partial}

\def\<{\langle}
\def\>{\rangle}

\def\lsim{\mathrel{\rlap{\lower3pt\hbox{\hskip0pt$\sim$}}
     \raise1pt\hbox{$<$}}}         
\def\gsim{\mathrel{\rlap{\lower4pt\hbox{\hskip1pt$\sim$}}
     \raise1pt\hbox{$>$}}}         

\def\lsim{\mathrel{\rlap{\lower3pt\hbox{\hskip0pt$\sim$}}
     \raise1pt\hbox{$<$}}}         
\def\gsim{\mathrel{\rlap{\lower4pt\hbox{\hskip1pt$\sim$}}
     \raise1pt\hbox{$>$}}}         

\usepackage{amsmath}
\usepackage{amsfonts}
\usepackage{verbatim}

\begin{document}

\begin{titlepage}

\begin{flushright}
{NYU-TH-05/18/11}

\today
\end{flushright}
\vskip 0.9cm

\centerline{\Large \bf Vainshtein Mechanism In $\Lambda_3$ - Theories}

\vskip 0.7cm
\centerline{\large Giga Chkareuli and David Pirtskhalava }

\vskip 0.3cm

\centerline{Center for Cosmology and Particle Physics,
Department of Physics,}
\centerline{\em New York University, New York,
NY, 10003, USA}

\centerline{}
\centerline{}

\vskip 1.cm

\begin{abstract}

We explore the space of spherically symmetric, static solutions in the decoupling limit of a class of non-linear covariant extensions of Fierz-Pauli massive gravity obtained recently in arXiv:1007.0443. In general, several such solutions with various asymptotic limits exist. We find their approximate short and long-distance behaviour and use numerical analysis to match them at the Vainshtein radius, $r_*$. Our findings indicate, that for a broad range of parameters, the theory does possess the Vainshtein mechanism, screening the scalar contribution to the gravitational force within $r_*$. In addition, there exists a class of solutions in the literature, for which the $1/r$ gravitational potential is completely screened within the Vainshtein scale. However, numerical analysis indicates, that for this type of solutions, the gravitational potential does not decay at spatial infinity.

\end{abstract}


\end{titlepage}

\section{Introduction and Summary}

It is a fundamental question of field theory, whether it is possible to write down a consistent, interacting lagranigian for a massive spin-2 state on Minkowski background. The first attempt to answer this question dates back to the original work by Fierz and Pauli \cite{FP}, who were able to find a unique ghost and tachyon - free mass term for a free Minkowski graviton. Breaking the gauge invariance of General Relativity (GR), the  model propagates five degrees of freedom of a four-dimensional spin-2 particle with a nonzero mass $m$, as required by Poincar\'e invariance.

The Fierz-Pauli (FP) model however can not describe the real world. The problem is connected with the observation of van Dam, Veltman and Zakharov (vDVZ) \cite{vdv, zakharov}, that even a tiny mass of the graviton which one would expect to be important only in deep infrared, leads to an $\mathcal{O}(1)$ modification of gravity at all scales. Indeed, in the $m\to 0$ limit, in addition to the usual GR interactions mediated by the helicity-2 component of the graviton, FP lagrangian describes the helicity-0 mode, coupled with gravitational strength to the trace of the external energy-momentum source,
\beq
\mathcal{L}_{m\to 0} \supset -\frac{1}{2•}(\p\pi)^2+\frac{1}{•M_P}\pi T.
\eeq 
Such a non-decoupling of the scalar degree of freedom would readily be ruled out by Solar System tests, which do not measure significant deviations from GR at the scales at hand.

As first noticed by Vainshtein \cite{Arkady} however (see also \cite{ddgv}), in a generic nonlinear completion of the FP model, the linear approximation breaks down at a large distance $r_*$ from an astrophysical source, such that the solar system tests should be compared to the predictions of the theory in a highly nonlinear regime. The effect originates in the strongly-coupled dynamics of the helicity-0 mode in a generic nonlinear massive gravity, which screens the scalar contribution to the gravitational potential. 

While restoring agreement with General Relativity,  nonlinear dynamics of the helicity-0 mode leads to yet more severe problems, connected with the appearance of a ghost on a generic background, as first found by Boulware and Deser (BD) in the Hamiltonian formalism \cite{BD}. There, the ghost manifests itself in the loss of the Hamiltonian constraint, leading to propagation of an additional, sixth ghost-like degree of freedom on top of the five of the linear Fierz-Pauli theory. 

An alternative method of detecting the BD ghost - first proposed in \cite{AGS}, is via the St\"uckelberg treatment of the theory. In this approach, one restores the broken gauge invariance of massive gravity by introducing four scalars, encoding the longitudinal degrees of freedom of a massive graviton at high energies. The high-energy regime is then captured by going to the "decoupling" limit, in which only the operators suppressed by the lowest scale are retained. Schematically, the decoupling limit lagrangian for the helicity-0 graviton on Minkowski background can be written in the following form for a generic extension of the FP model \cite{AGS,Deffayet,Creminelli},
\beq
\mathcal{L}_{m\to 0} \supset -\frac{1}{2•}(\p\pi)^2+\frac{1}{\Lambda^5_5}(\p^2 \pi)^3+\frac{1}{•M_P}\pi T.
\eeq
Here, $\Lambda_5=(M_P m^4)^{1/5}$ is the scale which is held finite in the limit at hand.  As noted above, the cubic self-interaction of the helicity-0 mode screens its contribution to the gravitational potential of a source of mass $M$ below the Vainshtein radius $r_*=(M/M_P^2m^4)^{1/5}$. Precisely due to this interaction however, the equation of motion for the scalar graviton is higher than quadratic in time derivatives, leading to the ill-posedness of the Cauchy problem and propagation of an extra scalar degree of freedom, which is inevitably a (BD) ghost. While infinitely heavy on Minkowski background, the ghost becomes light enough to destabilize any reasonable spherically symmetric solution in the theory - e.g. that of a localized lump of matter \cite{AGS, Deffayet, Creminelli}. 

Recently, a class of nonlinear generalizations of FP massive gravity has been proposed, which bypass these issues in the decoupling limit. By appropritely tuning the graviton potential, they cancel the dangerous $\p^2\pi$- dependent operators order-by order in nonlinearity \cite{drg1, drg2}. Moreover, the infinite number of contributions in the potential have been resummed into a two-parameter class of models, and the ghost-free property of the full theory has been proven up to and including the quartic order in nonlinearity in the Hamiltonian formalism \cite{drgt}. The cosmological solutions in the decoupling limit of this class of models have been studied in \cite{cosmo}. Interestingly, this type of extensions of FP massive gravity can be obtained in the recently proposed framework of theories with auxiliary extra dimensions \cite{g, dr} (for further studies of models with auxiliary dimensions, see \cite{hr, bm}).

In the present letter we will explore the space of spherically symmetric, static solutions in the Minkowski-space decoupling limit of the class of ghost-free extensions of the Fierz-Pauli model, found in \cite{drg1,drg2,drgt}. In this limit, one recovers a two-parameter scalar-tensor theory of a very special form, determined by symmetries at hand. We will call this class of models "$\Lambda_3$ - theories" below. The decoupling limit action is interesting in several ways; it represents a unique theory, to which any nonlinear, ghost-free massive gravity should reduce at high energies. Moreover, it is a certain scalar-tensor generalization (see \cite{drg2}) of higher-derivative, yet ghost-free scalar interactions, the simplest of which - the cubic one, was found in the decoupling limit of the DGP model \cite{dgp} in ref. \cite{lpr}. The cubic helicity-0 self-interaction of the DGP graviton has been extended to incorporate the full set of terms with similar properties - the so-called Galileons \cite{nrt}. Galileons are well known to enjoy special properties, such as second-order equations of motion, nonrenormalization of lowest-order interactions, presence of  the Vainshtein mechanism, etc. - all of these properties (including the Vainshtein mechanism, as we find below) are characteristic to the $\Lambda_3$ theories as well. Finally, the decoupling limit theory admits a ghost-free self-accelerated solution, at the same time providing a novel mechanism of decoupling the scalar degree of freedom from arbitrary matter \cite{cosmo}.

The study of an analogous problem has recently been conducted in \cite{ktn1,theo, ktn2}. It has been found by Koyama, Niz and Tasinato \cite{ktn2}, that for a class of parameters of the theory (described by   vanishing $\beta$ - a particular parameter, to be defined below), the theory admits the Vainshtein mechanism, successfully restoring agreement with GR to a very high precision at sub-Vainshtein scales. Moreover, for $\beta\neq 0$, a solution within the Vainshtein radius has been found in \cite{ktn2}, for which the $1/r$ contribution to the gravitational potential is completely screened, making it observationally unacceptable. In the present note we further study the space of spherically symmetric, static solutions in this class of theories and show that even if $\beta \neq 0$, for a broad subset of parameters (defined by $\beta<0$), another solution exists, which does exhibit the Vainshtein mechanism. Furthermore, we also find that the solution of \cite{ktn2} with the screened $1/r$ potential within the Vainshtein radius, matches a non-decaying solution outside.

The letter is organized as follows. In section 2 we describe the framework and some technical details, involved in the discussion. We derive the basic equations and comment on their similarity with those in Galileon theories. Section 3 deals with the discussion of spherically symmetric, static solutions. We find that in general, the theory admits multiple such solutions in the small and large distance regimes. Finally, we investigate their matching using numerical analysis.

\section{The Formalism and Technical Analysis}

Our starting point is the two-parameter decoupling limit action of the ghost-free class of generalizations of Fierz-Pauli theory \cite{drg1, drg2},
\beq
\mathcal{L}= -\frac{1}{2} h^{\mu\nu}\mathcal{E}_{\mu\nu}^{\alpha\beta} h_{\alpha\beta} + h^{\mu\nu}X^{(1)}_{\mu\nu}+\frac{\alpha}{\Lambda^{3}_{3}}h^{\mu\nu}X^{(2)}_{\mu\nu} +\frac{\beta}{\Lambda^{6}_{3}}h^{\mu\nu}X^{(3)}_{\mu\nu}+\frac{1}{ M_P}h^{\mu\nu}T_{\mu\nu}.
\label{lagr}
\eeq
Here, the first term represents the usual GR kinetic term for the
tensor mode, $\alpha$ and $\beta$ are two arbitrary
constants and $\Lambda_3=(M_P m^2)^{1/3}$ is the scale that is held fixed in the limit at hand; by $\(\mathcal{E}h\)\mn$  we denote the linearized Einstein
operator acting on metric perturbation, 
$\mathcal{E}^{\alpha\beta}_{\mu\nu} h_{\alpha\beta}=-\frac12 (\Box h\mn-
\p_\mu\p_\alpha h^\alpha_{\, \nu}-\p_\nu\p_\alpha h^\alpha_{\, \mu}+\p_\mu\p_\nu h
-\eta\mn \Box h + \eta\mn \p_\alpha \p_\beta h^{\alpha\beta} )$. 

The three identically conserved symmetric tensors  $X^{(n)}_{\mu\nu}[\Pi]$ depend on second
derivatives of the helicity-0 field $\Pi_{\mu\nu}\equiv \partial_\mu \partial_\nu \pi$ in the following way,

\begin{align}
X^{(1)}_{\mu\nu}=-\frac{1}{2}{\varepsilon_{\mu}}^{\alpha\rho\sigma}{{\varepsilon_\nu}^{\beta}}_{\rho\sigma}\Pi_{\alpha\beta}, \quad  \nonumber \\
X^{(2)}_{\mu\nu}=\frac{1}{2}{\varepsilon_{\mu}}^{\alpha\rho\gamma}{{\varepsilon_\nu}^{\beta\sigma}}_{\gamma}\Pi_{\alpha\beta}\Pi_{\rho\sigma}, \nonumber \\
X^{(3)}_{\mu\nu}={\varepsilon_{\mu}}^{\alpha\rho\gamma}{{\varepsilon_\nu}^{\beta\sigma\delta}}\Pi_{\alpha\beta}\Pi_{\rho\sigma}\Pi_{\gamma\delta}.
\nonumber
\end{align}
Being irrelevant for the present analysis, we ignore the helicity-1 mode of the massive graviton everywhere below; it enters the full nonlinear action at least quadratically, therefore allowing for a consistent solution for which it is not excited at all.

Before turning to the discussion of spherically symmetric solutions, let us briefly comment on symmetries of the action (\ref{lagr}). Under linearized diffeomorphisms, the helicity-2 graviton
transforms in the usual way, 
\beq
h_{\mn}\to h_{\mn}+\p_\mu\xi_\nu +\p_\nu\xi_\mu,
\eeq
while the scalar mode $\pi$ is invariant in the decoupling limit\footnote{This should not come as a surprise, since the helcity-0 field 
represents a \textit{physical} longitudinal polarization of the massive graviton in the decoupling limit.}. The diff invariance is then manifest in the fact that the currents $X^{(n)}_{\mn}$ are \textit{identically} conserved. On the other hand, there is also a manifest Galilean symmetry, under which the scalar mode shifts as 
\beq
\pi\to \pi+x^\mu b_\mu +c, \quad h_{\mn}=\textit{invariant}
\label{gt}
\eeq 
with $b_\mu$ and $c$ constants. Galilean transformations however can be redefined in a way, that realizes this symmetry only up to a total derivative, \cite{drg2}. Indeed, by making a helicity-2 redefinition, $h_{\mn}\to h'_{\mn}+\eta_{\mn}\pi+\alpha \p_\mu\pi\p_\nu\pi/\Lambda^3$, one can eliminate the $h X^{(1)}$ and $h X^{(2)}$ couplings in (\ref{lagr}), while the $h X^{(3)}$ coupling can not be eliminated\footnote{The $hX^{(1)}$ "coupling" is of course the usual conformal mixing of the helicity-2 and -0 modes, its elimination corresponding to transition to Einstein frame.}; as a result, one generates the full set of Galileon self-interactions for the scalar mode, as well as an additional coupling of the form  $\p_\mu\pi\p_\nu\pi T^{\mu\nu}$  on top of the usual coupling to the trace of the external energy-momentum source~\footnote{This coupling is irrelevant for a static, non-relativistic source. However it may play a significant role for gravitational lensing for instance, \cite{wyman}.}. Then, even if we redefine the transformations (\ref{gt}) so that $h'_{\mn}$ is invariant, the Galilean invariance will still persist up to a total derivative - precisely due to the Galileon structure of the decoupling limit theory in the Einstein frame. 

All of these facts indicate, that the spherically symmetric solutions in the given class of theories should not differ dramatically from those in modified gravities \'a la Galileon. In particular, for the subspace of parameters defined by $\beta=0$, for which the theory reduces to Galileon-modified gravity, this statement is automatic and massive gravity therefore possesses all well-known nice properties of Galileon field theories - the presence of Vainshtein mechanism among them. For $\beta\neq 0$ on the other hand, the situation is more subtle, and we will discuss this case in great detail below. 

Varying the action (\ref{lagr}) with respect to $h^{\mn}$ yields,
\beq
\mathcal{E}_{\mu\nu}^{\alpha\beta} h_{\alpha\beta}-X^{(1)}_{\mu\nu}-\frac{\alpha}{\Lambda^{3}_{3}}X^{(2)}_{\mu\nu}-\frac{\beta}{\Lambda^{6}_{3}}X^{(3)}_{\mu\nu}=\frac{1}{M_P•}T_{\mu\nu} ,
\label{eq01}
\eeq
while the equation of motion for the scalar mode reads as follows,
\beq
\partial_\alpha\partial_\beta h^{\mu\nu} (-\frac{1}{2}{\varepsilon_{\mu}}^{\alpha\rho\sigma}{{\varepsilon_\nu}^{\beta}}_{\rho\sigma}+ \frac{\alpha}{\Lambda_3^3}{\varepsilon_{\mu}}^{\alpha\rho\sigma}{{\varepsilon_\nu}^{\beta\gamma}}_{\sigma}\Pi_{\rho\gamma}+3\frac{\beta}{\Lambda_3^6}{\varepsilon_{\mu}}^{\alpha\rho\sigma}{{\varepsilon_\nu}^{\beta\gamma\delta}}\Pi_{\rho\gamma}\Pi_{\sigma\delta})=0.  \nn
\eeq
The latter equation can be rewritten in terms of the linearized curvature invariants in the following way, 
\ba
&R^{(1)}+2\frac{\alpha}{\Lambda^{3}_{3}}G^{(1) \mu\nu}\Pi_{\mu\nu}+\frac{6\beta}{\Lambda^{6}_{3}}\big (g^{\mu\nu}G^{(1)\alpha\beta}-g^{\mu\beta}G^{(1)\alpha\nu}+g^{\alpha\beta}R^{(1)\mu\nu}-g^{\alpha\nu}G^{(1)\mu\beta}
\nn \\ &-R^{(1)\mu\alpha\nu\beta}\big )\Pi_{\mu\nu}\Pi_{\alpha\beta}=0,
\label{eq02}
\ea
where $R^{(1)}_{\mn}$, $G^{(1)}_{\mn}\equiv \mathcal{E}_{\mu\nu}^{\alpha\beta} h_{\alpha\beta}$ and $R^{(1)}_{\mn\alpha\beta}$ denote respectively the linearized Ricci, Einstein and Riemann tensors composed of the helicity-2 field $h_{\mn}$. Presence of the Riemann tensor in the latter expression makes it impossible in general to algebraically reduce the system of equations of motion to a set of pure - $\pi$ and pure - $h_{\mn}$ equations, whereas for $\beta=0$ this is possible. This is just another manifestation of the fact found in \cite{drg2}, that one can eliminate the $h^{\mu\nu}X^{(1,2)}_{\mu\nu}$ couplings by a nonlinear redefinition of the tensor mode, while the coupling $h^{\mu\nu}X^{(3)}_{\mu\nu}$, if present, can not be absorbed by any such redefinition of variables. 

The most general spherically symmetric ansatz for the metric perturbation is given as follows,
\begin{equation}
h_{00}=a(r), ~~~h_{ij}=f(r)\delta_{ij}+b(r)n_{i}n_{j},
\label{tensoranzats}
\end{equation}
where $a,f$ and $b$ are general functions of the radial variable, while $n_i$ denotes the unit vector in the radial direction. We will exploit the diff invariance of (\ref{lagr}) to set $b(r)=0$. Substituting the ansatz into the 0-0 component of the Einstein's equation (\ref{eq01}), one obtains,
\beq
\frac{1}{r^2}\left(-r^2f^{\prime}+r^{2}\pi^{\prime}-\frac{\alpha}{\Lambda^{3}_{3}}r(\pi^{\prime})^{2}-\frac{2\beta}{\Lambda^{6}_{3}}(\pi^{\prime})^{3}\right)^{\prime}=\frac{1}{M_P} T_{00},
\label{eq00}
\eeq
where a prime denotes a derivative by $r$. The spatial components of (\ref{eq01}) on the other hand, reduce to the following two equations,
\beq
&\left(rf^{\prime}-ra^{\prime}-2r\pi^{\prime}+\frac{\alpha}{\Lambda^{3}_{3}}(\pi^{\prime})^{2}\right)^{\prime}=0, \nn
\\&\left(\frac{f^{\prime}}{r}-\frac{a^{\prime}}{r}-\frac{2\pi^{\prime}}{r}+\frac{\alpha}{\Lambda^{3}_{3}}\frac{(\pi^{\prime})^{2}}{r^{2}}\right)^{\prime}=0. \nonumber
\eeq
These equations are consistent only if the integration constants vanish, leading to a single equation,
\beq
rf^{\prime}-ra^{\prime}-2r\pi^{\prime}+\frac{\alpha}{\Lambda^{3}_{3}}(\pi^{\prime})^{2}=0.
\label{eqij}
\eeq
Finally, the equation of motion for the helicity-0 field $\pi$ is of the form $\p_\mu J^{\mu}_\pi=(r^2 J^r_\pi)'/r^2=0$, where $J^{\mu}_\pi$ denotes the conserved Noether current of the shift symmetry $\pi\to\pi+c$. Since the equations of motion that follow from the lagrangian (\ref{lagr}) contain no more than two derivatives per field,  $J^\mu_\pi$ can not contain higher than a single derivative of $\pi$, leading to an algebraic equation for $\pi'$, just as in the case of Galileons \cite{nrt}. Plugging the anzats (\ref{tensoranzats}) into eq. (\ref{eq02}), the scalar equation reads, 
\beq
\left(2r^2f^{\prime}-r^2a^{\prime}+2\frac{\alpha}{\Lambda^{3}_{3}}r\pi^{\prime}\left(a^{\prime}-f^{\prime}\right)+6\frac{\beta}{\Lambda^{6}_{3}}a^{\prime}(\pi^{\prime})^2\right )^{\prime}=0.
\label{eqpi}
\eeq

Below we explore the space of spherically symmetric, static solutions to the system (\ref{eq00})-(\ref{eqpi}). We will be interested in a spherical source of size $R$, constant density $\rho$ and negligible pressure, so that the only nonvanishing component of the energy-momentum tensor is $T_{00}=\rho~ \theta(R)$. Integrating eqs. (\ref{eq00}) and (\ref{eqpi}) subject to boundary conditions $a(0)=f(0)=\pi(0)=0$, one reduces the system to the following set of equations for the radial derivatives of the fields,
\beq
a^{\prime}=-\frac{M(r)}{M_{p}r^2}-\Lambda_3^3 r \lambda (1+2\beta \lambda^2),
\label{1}
\eeq
\beq
f^{\prime}=-\frac{M(r)}{M_{p}r^2}+\Lambda_3^3 r\lambda (1-\alpha\lambda-2\beta\lambda^2),
\label{2}
\eeq
\beq
3\lambda-6\alpha\lambda^2+(2\alpha^2-8\beta)\lambda^3-12\beta^2\lambda^5= \left\{ \begin{array}{ll}
(\frac{r_*}{r•})^3(1+6\beta\lambda^2) & \mbox{Outside the source} \\ (\frac{r_*}{R})^3(1+6\beta\lambda^2) & \mbox{Inside the source} \\ \end{array} 
\right .
\label{3}
\eeq
where $M(r)$ denotes the mass of the source within a sphere of radius $r$, and we have defined,
\beq
\lambda\equiv \frac{\pi'}{\Lambda^3_3 r}, \qquad r_*\equiv \( \frac{M}{M^2_P m^2•} \)^{1/3}.
\eeq

\section{Solutions}

We will first concentrate on solutions to the system  (\ref{1})-(\ref{3}) outside the source. The helicity-0 equation (\ref{3}) is quintic in $\lambda$; it can not be solved exactly and in general we should expect multiple solutions. Outside the source, there exists a distance scale - the Vainshtein radius $r_*$, at which a typical solution changes regime; we will therefore divide the space further into regions outside and inside the Vainshtein radius and solve the system (\ref{1})-(\ref{3}) in each of these regions separately. The nontrivial part is to find an appropriate matching of solutions in the two regions, which in principle requires an exact solution to the system. Having no handle on exact analytic solutions, we will implement numerical analysis in support of our findings.

\subsubsection*{3.1 Solutions beyond the Vainshtein radius}

For $r\gg r_*$, there exist two types of solutions. One of them (which we call "Solution 1" below) is obtained by simply neglecting all nonlinearities in $\lambda$ in eq (\ref{3}), so that
\beq
\lambda=\frac{1}{3•}\Big (\frac{r_*}{r}\Big )^3\Rightarrow \pi=-\frac{1}{3}\frac{M}{M_Pr}, \quad a=\frac{4}{•3}\frac{M}{M_P r•}, \quad f=\frac{2}{•3}\frac{M}{M_P r•},
\eeq
with corrections suppressed by powers of $r_*/r$. This is an asymptotically flat solution of the linear FP theory - fields vanish at infinity, and it manifestly exhibits the vDVZ discontinuity.  This type of asymtpotic behaviour one encounters in straightforward nonlinear generalizations of the FP masisve gravity, see e.g. \cite{dkp,bdz1,bdz2,bdz3}. 

Since the right hand side of (\ref{3}) contains a small overall factor $(r_*/r)^3$, formally a second solution ("Solution 2") is possible, for which the left hand side vanishes in the zeroth approximation, 
\beq
\lambda\propto \pi'/r=C \Rightarrow \pi \sim \Lambda_3^3 r^2, \quad a,f \sim \Lambda_3^3 r^2,
\eeq
with $C$ a nonzero constant and up to corrections suppressed by powers of $r_*/r$. This is an asymptotically non-decaying solution - both the helicity-0 and helicity-2 fields increase with distance on this branch (see also the discussion in section 3.3); the fact of its existence will play an important role below. 
 
 \subsubsection*{3.2 Solutions within the Vainshtein radius}
 
Well within the Vainshtein radius, $r_*/r\gg1$, and for $\beta<0$, there exists a solution ("Solution 3") for which the r.h.s. of eq. (\ref{3}) vanishes in the zeroth approximation. On this branch, the leading $r$-depencence of the fields reads as follows,  
 \beq
 \lambda=\sqrt{-\frac{1}{6\beta}}+\mathcal{O}\( \(\frac{r}{r_*}\)^3\) \Rightarrow \pi\sim \Lambda_3^3 r^2 ,~~ a=f=\frac{M}{M_P r}+\mathcal{O} \( \( \frac{r}{r_*} \)^3 \frac{M}{M_P r•}\),
 \label{sol3}
 \eeq
and one recovers with great precision the gravitational potential of GR, up to tiny corrections due to the helicity-0 graviton. One should note, that these corrections are smaller than analogous ones in the DGP model (see \cite{dgz,ls}). For $\beta<0$ therefore, the theory admits the Vainshtein mechanism! 

Finally, there exists an interesting solution found by Koyama, Tasinato and Niz ("Solution 4") \cite{ktn1, ktn2}, for which the usual $1/r$ part of the gravitational potential is completely screened inside the Vainshtein radius. Mathematically, this corresponds to last terms on each side of eq. (\ref{3}) being dominant over the rest of the contributions.  Then, the leading dependence on $r$ in $a$ and $f$ cancels, leaving the following expressions for the solution at hand,  
\beq
\lambda\simeq \(-\frac{1•}{2\beta•}\)^{1/3}\frac{r_*}{r•},\quad a\sim \frac{M}{M_Pr^2_*•} r , \quad f \sim \frac{M}{M_Pr_*•} \ln r .
\eeq
This modifies the gravitational potential within $r_*$ to an observationally unacceptable form.

Within the source, as it follows from (\ref{3}), $\lambda$ is constant, matching the value at $r=R$ of a particular solution inside the Vainshtein radius discussed above (Solution 3 or 4). This means, that $\pi$ and its first derivative are continuous across the surface of the source.

\subsubsection*{3.3 Matching at the Vainshtein radius}

Matching the solutions found inside and outside the Vainshtein radius is perhaps the most nontrivial task to perform. As remarked above, this procedure can not be carried out analytically and numerical analysis should be addressed. We therefore will not be able to claim full generality in this section - we will rather present results for a particular choice of the parameters $\alpha$ and $\beta$; these results seem pretty generic - at least to the extent that numerical analysis can provide. Our findings indicate, that if $\beta<0$, Vainshtein mechanism works (at least for all values of the parameters consistent with this condition, that we have been able to check). This means that the GR solution (Solution 3) which exhibits the Vainshtein mechanism inside $r_*$, does match the asymptotically flat Solution 1, exhibiting the vDVZ discontinuity beyond the Vainshtein scale. One example of this is shown in fig.1, which gives the dependence of $\lambda$ on $r/r_*$ for $\alpha=-1$ and $\beta=-1$. The asymptotics of the plotted solution at $r/r_*\to 0$ and $r/r_*\to\infty$ clearly coincide with the Solution 3 ( $\lambda\simeq \textit{const}$) and Solution 1 ($\lambda\propto 1/r^3$) respectively. On the other hand, the asymptotically non-decaying Solution 2 ($\lambda\simeq \textit{const}$) outside the Vainshtein radius matches the Solution 4 ($\lambda\propto 1/r$) inside, as shown on an analogous plot in fig. 2 for the same values of the parameters. For $\beta>0$, only the latter type of solutions exists (see fig. 3).

It is important to address the fact, that the leading terms in the expressions for $\lambda$ for Solns. 2 and 3 do not depend on the mass of the source. One can then ask the question: how physical  are these solutions, since the fields do not vanish even in the $M\to 0$ limit?  In this limit, the non-decaying Solution 2 describes an asymptotically non-flat configuration with nonzero (and increasing with distance) gravitational potential throughout the whole space. On the other hand, the region of validity of the the sub-Vainshtein Solution 3, $r\ll r_*$, shrinks to zero for the vanishing mass of the source. Moreover, for any nonzero $M$, corrections to the gravitational potential are tiny within the Vainshtein radius, as noted above
\footnote{In fact, these arguments are valid only if one does not consider quantum corrections. In the full quantum theory, one can make predictions only down to distances of order of the inverse cutoff of the effective theory, $r_q \sim \Lambda_3^{-1}$.  For Planckian sources, $r_q$ merges with the Vainshtein radius $r_*$  and the theory transitions from the linear regime straight into the quantum one, in which definite predictions can not be made. For any transplanckian source $M\gg M_P$ on the other hand, the Solution 3 describes a well defined region in which the helicity-0 contribution is screened and one recovers GR with great precision, yet quantum corrections are under control. } . Finally, for $M=0$, there of course exists a physical solution, on which all fields are identically zero.
  
A comment on yet another class of solutions on asymptotically non-flat backgrounds may be in place here; on a self-accelerated background of (\ref{lagr}) , the $hX^{(1)}$ mixing vanishes (while the kinetic term for the helicity-0 graviton comes from the curvature of the backround spacetime), making it possible for the scalar mode $\pi$ not to be excited at all by a localized source, thus decoupling it from arbitrary matter (for details, see \cite{cosmo}).   
 
\begin{figure}[htb]
\begin{center}
\includegraphics[height=2.5in,width=4in,angle=0]{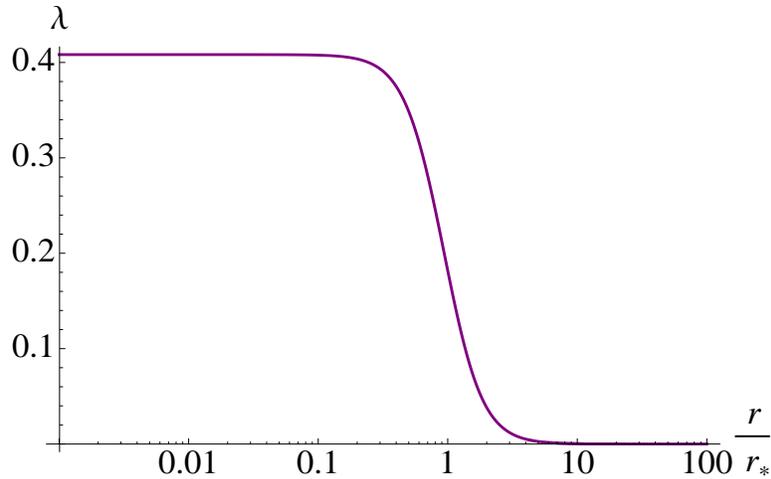}
\caption{ The solution exhibiting the Vainshtein mechanism for $\alpha=-1$ and $\beta=-1$.  }
\end{center}
\end{figure}

\begin{figure}[htb]
\begin{center}
\includegraphics[height=2.5in,width=4in,angle=0]{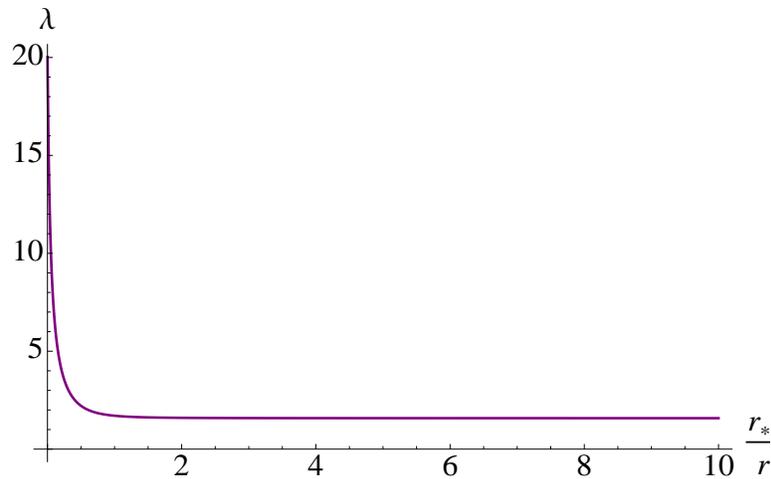}
\caption{ The asymptotically non-decaying solution for $\alpha=-1$ and $\beta=-1$. }
\end{center}
\end{figure}

\begin{figure}[htb]
\begin{center}
\includegraphics[height=2.5in,width=4in,angle=0]{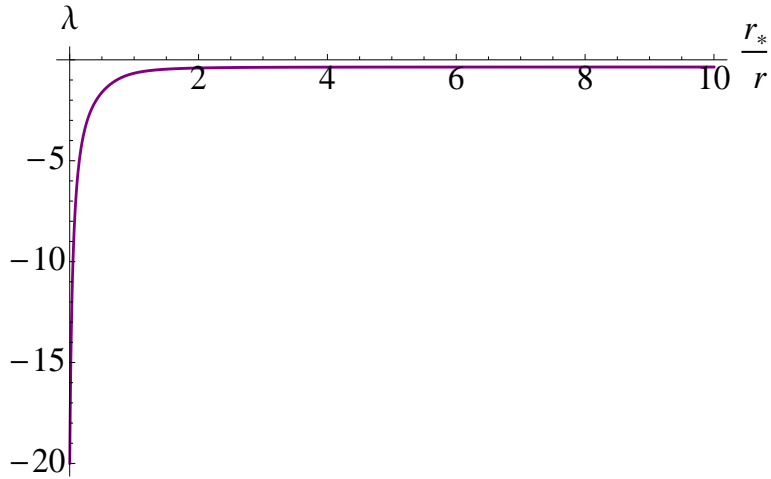}
\caption{ The asymptotically non-decaying solution for $\alpha=-1$ and $\beta=1$. No asymptotically decaying solutions exist for this choice of parameters.  }
\end{center}
\end{figure}

In summary, our findings indicate that in general there are multiple spherically symmetric, static solutions in the Minkowski-space decoupling limit of ghost-free nonlinear extensions of the Fierz-Pauli model, found in \cite{drg1,drg2,drgt}. Moreover, for a wide range of parameters, defined by $\beta<0$, the Vainshtein mechanism seems to work in the theory - there exists an asymptotically-flat solution, which screens the contribution of the helicity-0 mode at sub-Vainshtein scales, successfully hiding it from the Solar System tests.
In addition, there exists an asymptotically non-decaying solution, which totally screens the $1/r$ potential inside the Vainshtein radius \cite{ktn2}, dramatically reducing the gravitational attraction at these scales - this scenario is clearly ruled out by observations.

\vspace{10pt}

\noindent {\bf Acknowledgements:}
We would like to thank Gregory Gabadadze for guidance, support and valuable discussions throughout the completion of the work. We would also like to thank Lasha Berezhiani for useful discussions and  Kazuya Koyama and Gustavo Niz for valuable correspondence and comments on the manuscript. G.C. and D.P are respectively supported by the MacCracken and Mark Leslie graduate assistantships at New York University.

\end {document}